\documentclass[cameraready]{Interspeech}
\usepackage{booktabs}
\usepackage{tikz, pgfplots}
\usepackage{xcolor} 
\usepackage{pgfplots}
\pgfplotsset{compat=1.18}  
\usepackage{enumitem}

\definecolor{colorSpeech}{HTML}{1f77b4}      
\definecolor{colorEntertain}{HTML}{ff7f0e}   
\definecolor{colorEducation}{HTML}{2ca02c}   
\definecolor{colorMusic}{HTML}{d62728}       
\definecolor{colorAnnounce}{HTML}{9467bd}    
\definecolor{colorMedia}{HTML}{8c564b}       
\definecolor{colorEmergency}{HTML}{e377c2}   
\definecolor{colorCultural}{HTML}{bcbd22}    
\definecolor{colorOthers}{HTML}{7f7f7f}      
\title{TW-Sound580K: A Regional Audio–Text Dataset with Verification-Guided Curation for Localized Audio-Language Modeling}

\author[affiliation={1,2}]{Hao-Hui }{Xie}
\author[affiliation={1}]{Ho-Lam }{Chung}
\author[affiliation={1},orcid=0009-0007-3994-6433]{Yi-Cheng }{Lin}
\author[affiliation={1}]{Ke-Han}{Lu}
\author[affiliation={1}]{Wen-Ze}{Ren}
\author[affiliation={2}]{Xie}{Chen}
\author[affiliation={1}, correspondingauthor]{Hung-yi}{Lee}

\address{
  $^1$ National Taiwan University \\
  $^2$ Shanghai Jiao Tong University
}

\email{xhh666@sjtu.edu.cn}

\begin{document}
\maketitle

\begin{abstract}
Large Audio-Language Models (LALMs) typically struggle with localized dialectal prosody due to the scarcity of specialized corpora. We present TW-Sound580K, a Taiwanese audio-text instruction dataset developed through a Verify-Generate-Critique (VGC) protocol. This pipeline leverages Dual-ASR validation to filter 522K raw clips, subsequently expanding them into 580,000 high-fidelity instruction pairs using a teacher model. The dataset’s utility is demonstrated through Tai-LALM, which fine-tunes a DeSTA 2.5-Audio-initialized backbone and incorporates a dynamic Dual-ASR Arbitration strategy to optimize transcription selection during inference. On the TAU Benchmark, Tai-LALM reaches 49.1\% accuracy, marking a 6.5\% absolute improvement over the zero-shot baseline (42.6\% with ASR text conditioning). This confirms that integrating regional corpora with rigorous curation and dynamic arbitration significantly enhances LALM performance on localized speech.
\end{abstract}

\section{Introduction}
Recent advancements in Large Audio-Language Models (LALMs) \cite{rubenstein2023audiopalm, huang2025dynamicsuperb} have improved multimodal reasoning across various speech and environmental contexts \cite{gong2023ltu, tang2023salmonn, chu2023qwenaudio}. Despite this progress, models often underperform in culturally specific regions due to a localization gap \cite{cao2024culturellm}. In linguistically diverse areas like Taiwan, audio comprehension relies on distinct acoustic markers, primarily non-standard dialectal prosody and regional environmental soundmarks. Current models frequently treat these nuanced signals as out-of-distribution noise \cite{shor2022universal, yang2024taiwanese}. The scarcity of localized training data leaves these regional acoustic features under-represented. As a result, models struggle to decode these patterns and are prone to acoustic hallucinations \cite{kuan2024understanding}, such as forcibly transcribing environmental sounds into nonsensical text. This highlights the difficulty of aligning regional acoustics with general cross-cultural semantics \cite{cao2024culturellm}. Addressing this issue requires high-fidelity, region-specific data rather than relying solely on generic semantic recognition \cite{yang2024taiwanese, lin2025mitigating}.

To mitigate this data scarcity, we introduce TW-Sound580K, a large-scale Taiwanese audio-text instruction dataset. Built upon approximately 522,000 raw audio clips , the dataset is expanded to over 580,000 diverse instruction-response pairs via a teacher LLM. This corpus is specifically designed to capture the local ``acoustic long-tail'' defined in this work as the sparse and imbalanced distribution of region-specific environmental sounds and minor dialectal variants. It provides extensive coverage of both regional dialects and unique regional soundmarks.

Constructing clean supervision from such culturally dense data presents a practical challenge. Standard Automatic Speech Recognition (ASR) systems typically fail to process non-lexical environmental cues, yet entirely bypassing ASR compromises the transcription accuracy of complex dialects. To build a high-quality dataset without introducing hallucination risks, we implement a Verify-Generate-Critique (VGC) curation pipeline integrated with a Dual-ASR filtering strategy. This mechanism uses heterogeneous ASR systems to validate speech data and filter transcription inconsistencies. For environmental audio, it acts as a speech-absence verifier, relying on a teacher model's critique to ensure data purity.

Additionally, to properly evaluate models trained on this corpus, we propose a dynamic Dual-ASR Arbitration mechanism during the inference stage. Guided by acoustic-conditioned perplexity (AC-PPL), this arbiter dynamically selects the most accurate transcription from the ASR outputs. This effectively reduces the risk of run-time hallucinations when encountering heavy dialectal noise or nuanced regional acoustics.

The main contributions of this work are summarized as follows:
\begin{itemize}
    \item \textbf{The TW-Sound580K Dataset:} We introduce a large-scale instruction-tuning corpus targeting the Taiwanese acoustic long-tail, expanded from 522K raw audio clips. It provides high-quality supervision for regional dialects and local soundmarks. Due to copyright and licensing constraints, the raw audio data cannot be directly distributed. However, to ensure reproducibility while adhering to double-blind review policies, the source URLs, metadata, and associated crawler scripts are prepared for public release upon de-anonymization.
    
    \item \textbf{Automated Curation Pipeline and Dynamic Inference Arbitration:} We design a VGC-based data processing pipeline integrated with Dual-ASR filtering to ensure high-fidelity dataset construction. Furthermore, we propose an AC-PPL-guided dynamic arbitration strategy during inference to dynamically select the most accurate transcription, effectively reducing the risk of run-time hallucinations.
    
    \item \textbf{Empirical Validation via Tai-LALM:} We fine-tune Tai-LALM (initialized with DeSTA 2.5-Audio weights) to validate the corpus and inference strategy. Evaluated on the TAU Benchmark, the model achieves 49.1\% accuracy, outperforming the zero-shot baseline (42.6\%) by 6.5\% and a naive SFT baseline trained on unfiltered data by 2.7\%.
\end{itemize}
Ultimately, this work provides not just a localized corpus, but a reproducible framework from data curation to inference arbitration that effectively bridges the localization gap for regional audio understanding in LALMs.
\begin{figure*}[t]
    \centering
    \vspace{-0.8cm} 
    
    \includegraphics[width=1.0\textwidth]{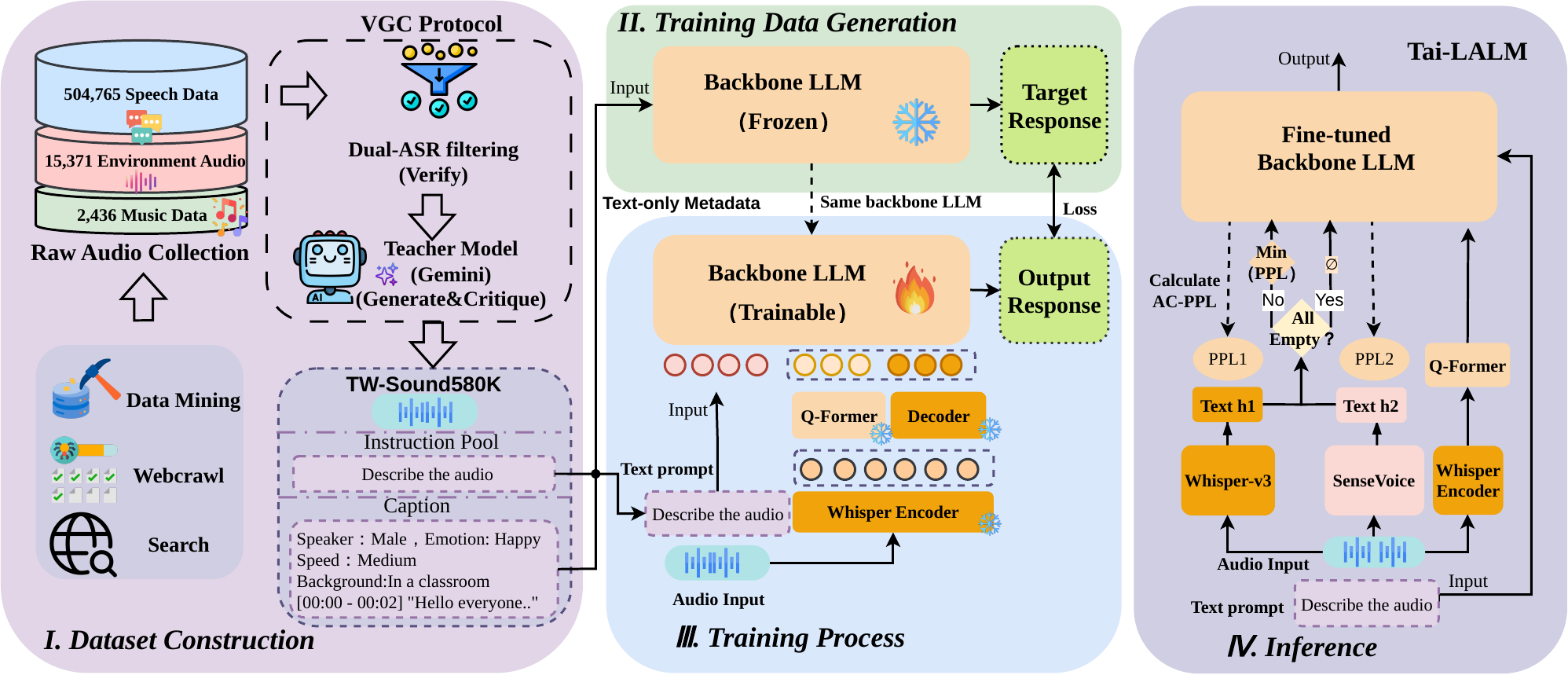}
    
    \vspace{-0.2cm} 
    
    \caption{The proposed framework for TW-Sound580K dataset construction and Tai-LALM fine-tuning, illustrating the DeSTA 2.5-Audio-based localization pipeline.}
    \label{fig:overview}
    
    \vspace{-0.3cm} 
\end{figure*}
\section{Background}

Foundational corpora like AudioSet \cite{gemmeke2017audioset} and LibriSpeech \cite{panayotov2015librispeech} predominantly feature standard acoustic environments and dominant accents. Similarly, large-scale Mandarin datasets such as WenetSpeech \cite{zhang2022wenetspeech} favor Standard Mandarin, effectively marginalizing regional prosody and dialectal variations. While modern instruction-tuning sets \cite{kreuk2023audiogen, tang2023salmonn} facilitate multimodal reasoning, they focus on cross-cultural semantics \cite{chu2025when} rather than the “acoustic long-tail” of local soundmarks \cite{kim2025wow}. Regional datasets like KeSpeech \cite{nie2022kespeech}, though specialized, remain confined to basic ASR tasks and lack the complex reasoning pairs necessary for LALMs. These limitations create a distinct localization gap \cite{cao2024culturellm} that the TW-Sound580K dataset aims to bridge.

Training LALMs on in-the-wild regional audio poses a distinct curation challenge \cite{ding2025less}. The structural complexity of culturally rich data often clashes with standard SFT paradigms \cite{ouyang2022training}, leading to noisy labels and semantic drift \cite{udandarao2025spe-langy}. In these cases, strong LLM priors may generate plausible but hallucinated text divorced from the auditory input \cite{hsu2025reducing,beyond2025asr}. Without effective modality-aware filtration, model performance on localized tasks degrades \cite{lin2025mitigating}, necessitating a refinement pipeline like the Verify-Generate-Critique (VGC) mechanism to ensure supervision quality.

Architectural constraints during inference further impede localization. Cascaded systems ensure lexical accuracy but discard paralinguistic cues \cite{shor2022universal, tau2025benchmark}, while E2E models preserve fidelity but struggle with regional phonetic instability \cite{cao2024culturellm}. Hybrid architectures like DeSTA 2.5-Audio \cite{desta2025} attempt to bridge these pathways but still face the ``premature commitment'' problem \cite{sperber2019attention}. For complex regional dialects, a single ASR system often exhibits high transcription errors, mapping acoustic uncertainty into erroneous text and triggering hallucinations \cite{atwany2025lost, kuan2025teaching, frieske2024hallucinations}. This highlights the need for a dynamic arbitration mechanism to select optimal transcriptions and overcome single-system performance bottlenecks.


\section{Methodology}

To bridge the localization gap, our data-centric pipeline is structured into four key stages: (I) Dataset Construction, (II) Training Data Generation, (III) the multimodal Training Process, and (IV) Inference featuring a dynamic Dual-ASR Arbitration mechanism.


\subsection{TW-Sound580K: Socio-Functional Data Engineering for Taiwan}
To mitigate representational sparsity in regional audio modeling, our dual-ASR pipeline filters 522,572 raw independent audio clips from Taiwan-centric sources into a refined collection of 456,832 validated samples, totaling approximately 3,536.78 hours. Based on this high-fidelity corpus, we utilize a teacher LLM to generate one or more instructions per audio clip, expanding the dataset into 580,000 diverse audio-text pairs, hereafter referred to as TW-Sound580K.

To ensure scalable annotation, we employ Qwen3 to perform multi-label socio-functional classification based on the teacher-generated semantic descriptions. Although speech-inclusive clips dominate the dataset, the \textit{Conversation} tag comprises only 46.4\% of the total assigned labels. The remaining 53.6\% specifically targets the Taiwanese acoustic long-tail, capturing unique dialectal prosody and localized soundmarks. This distribution compels the model to process Taiwanese cultural acoustic cues as integral semantic features rather than background noise.

\begin{figure}[htbp]
    \centering
    \begin{tikzpicture}
        \definecolor{colConversation}{HTML}{C0DF36}
        \definecolor{colEntertain}{HTML}{7FCE53}
        \definecolor{colEducation}{HTML}{4CBB70}
        \definecolor{colMusic}{HTML}{26A67B}
        \definecolor{colOthers}{HTML}{239587}
        \definecolor{colAnnounce}{HTML}{23838D}
        \definecolor{colMedia}{HTML}{2B6F92}
        \definecolor{colEmergency}{HTML}{2E588F}
        \definecolor{colCultural}{HTML}{373F7C}

        \begin{axis}[
            ybar,
            width=\linewidth,  
            height=5cm,        
            ymin=0, ymax=55,   
            ymajorgrids=true,
            grid style={dashed, gray!40, line width=0.6pt},
            axis on top=false,
            symbolic x coords={Conversation, Entertainment, Education, Music, Others, Announcement, Media, Emergency, Cultural},
            xtick={Conversation, Entertainment, Education, Music, Others, Announcement, Media, Emergency, Cultural},
            xticklabel style={rotate=30, anchor=north east, font=\footnotesize, inner sep=2pt},
            yticklabel style={font=\footnotesize}, 
            nodes near coords,
            nodes near coords align={vertical},
            every node near coord/.append style={font=\footnotesize, yshift=1pt}, 
            axis x line=bottom,
            axis y line=left,
            x axis line style={black, thick},
            y axis line style={black, thick},
            bar width=14pt,        
            bar shift=0pt,        
            enlarge x limits=0.08, 
            /pgf/number format/fixed,
            /pgf/number format/precision=1,
            point meta=rawy,
        ]
            \addplot[fill=colConversation, draw=none] coordinates {(Conversation, 46.4)};
            \addplot[fill=colEntertain, draw=none] coordinates {(Entertainment, 17.1)};
            \addplot[fill=colEducation, draw=none] coordinates {(Education, 16.5)};
            \addplot[fill=colMusic, draw=none] coordinates {(Music, 12.4)};
            \addplot[fill=colOthers, draw=none] coordinates {(Others, 2.7)};
            \addplot[fill=colAnnounce, draw=none] coordinates {(Announcement, 2.0)};
            \addplot[fill=colMedia, draw=none] coordinates {(Media, 1.4)};
            \addplot[fill=colEmergency, draw=none] coordinates {(Emergency, 0.8)};
            \addplot[fill=colCultural, draw=none] coordinates {(Cultural, 0.7)};
        \end{axis}
    \end{tikzpicture}
    \caption{Label occurrence distribution in the TW-Sound580K dataset.}
    \label{fig:vertical_bar_chart}
\end{figure}

\subsection{The VGC Protocol: Modality-Aware Filtration}
To extract high-fidelity supervision and avoid severe semantic hallucinations, we introduce the Verify-Generate-Critique (VGC) protocol:

\begin{enumerate}[leftmargin=*]
    \item Verify (Conditional Routing): We procure transcriptions from two heterogeneous ASR engines to compute a semantic consistency score $\mathcal{S}$ (based on text similarity). To preserve speech-free soundmarks, clips where both ASRs yield empty outputs bypass the text check. Conversely, speech samples with $\mathcal{S}$ below a predefined empirical threshold $\tau$ are explicitly pruned to prevent irrecoverable dialectal noise ingestion.
    
    \item Generate (Acoustic-Constrained Distillation): A powerful native-audio Large Language Model  acts as our Teacher Model. By processing raw continuous audio without referring to validated ASR transcriptions, restrictive zero-shot prompting constrains outputs to verifiable paralinguistic and environmental features, preventing cross-modal hallucinations.
    
    \item Critique (Self-Reflective Audit): The teacher model conducts a secondary review to prune any ungrounded descriptors from the captions. This process ensures that the Taiwan-centric instruction data is strictly anchored to actual acoustic cues while preserving the full original audio collection.
\end{enumerate}

\subsection{Inference-Time Perceptual Arbitration}
To mitigate errors from incorrect ASR injections, we employ the SFT-tuned Tai-LALM as a real-time Perceptual Arbiter. Given a set of candidate transcriptions $H$ from heterogeneous ASR engines, the arbiter selects the optimal transcription $\hat{h}$ by minimizing the Acoustically-Conditioned Perplexity (AC-PPL):

\begin{equation}
\hat{h} = \arg \min_{h \in H} \exp \left( - \frac{1}{|h|} \sum_{i=1}^{|h|} \log P(w_i \mid w_{<i}, \mathbf{z}_A; \theta) \right)
\label{eq:ppl_selection}
\end{equation}

where $\mathbf{z}_A$ is the latent acoustic representation and $|h|$ is the candidate's sequence length. We adopt AC-PPL because it measures how well a candidate aligns with the model's internal understanding. Crucially, if all candidates in $H$ are detected as empty soundmarks, the arbiter bypasses text injection and shifts to strictly unconditioned pure-audio reasoning.

\subsection{Implementation and Training Objectives}
Following DeSTA 2.5-Audio, we employ a self-generated target mechanism. First, a frozen, text-only backbone LLM processes VGC-curated metadata to generate target responses $Y$. Second, an identical multimodal backbone, bridged to a frozen Whisper Encoder via a Q-Former, is fine-tuned. Low-Rank Adaptation (LoRA) is applied exclusively to the backbone's attention layers ($\phi$). 

During training, the model is optimized to minimize the autoregressive loss by conditioning on both the continuous acoustic representation $\text{Q}(\mathbf{z}_A)$ and the text generated by the built-in ASR ($h_{gt}$):
$$
\mathcal{L}_{\text{SFT}} = - \sum_{t=1}^{T} \log P(y_t \mid y_{<t}, h_{gt}, \text{Q}(\mathbf{z}_A); \phi)
$$




\section{Experiments}
\label{sec:experiments}

\subsection{Experimental Setup}

\textbf{Implementation Details:} The proposed model, Tai-LALM, is developed as a localized adaptation of the DeSTA 2.5-Audio framework. It inherits its architectural configuration and pre-trained weights directly from the DeSTA 2.5-Audio finetune stage, utilizing the Llama-3-8B-Instruct backbone. Modality alignment is facilitated by a pre-trained 64-query Q-Former. 

To evaluate the absolute performance gains from regional supervision, we employ the zero-shot DeSTA 2.5-Audio as our primary baseline. Tai-LALM is fine-tuned via LoRA ($r=32, \alpha=32$) exclusively on the backbone's attention layers. The VGC protocol utilizes Whisper-v3 \cite{radford2023whisper} and SenseVoice \cite{sensevoice2024} as heterogeneous ASR engines (threshold $\tau = 0.6$), with Gemini-2.5-Pro \cite{gemini25_pro2025} providing acoustic-constrained distillation. Training is conducted on a single NVIDIA GH200 GPU for 2 epochs with a global batch size of 32 and a learning rate of $5 \times 10^{-5}$.

\textbf{Negative Control :} To isolate the impact of our VGC  and inference-time perceptual arbitration, we fine-tune the DeSTA 2.5-Audio backbone directly on the Unfiltered Raw TW-Sound580K+ without inference arbitration. While the extensive scale of this raw dataset  inherently provides a strong performance floor, residual transcription noise caps the model's potential to capture fine-grained regional nuances. This demonstrates that scale alone cannot fully bridge the acoustic gap.

\subsection{Main Results on TAU Benchmark}

We evaluate auditory reasoning on the TAU Benchmark \cite{tau2025benchmark} (1,794 queries), which spans single-hop and multi-hop tasks. Table~\ref{tab:tau_main} compares Tai-LALM against state-of-the-art systems.

\begin{table}[htbp]
\centering
\caption{Performance on the TAU Benchmark. Tai-LALM significantly outperforms zero-shot baselines and the naive SFT control.} 
\label{tab:tau_main} 
\resizebox{0.95\linewidth}{!}{ 
\begin{tabular}{lcccc}
\toprule
\textbf{System} & \textbf{Params} & \textbf{Single} & \textbf{Multi} & \textbf{Overall} \\ \midrule
Gemini 2.5 Pro (Teacher) & -- & 72.4 & 73.9 & 73.0 \\
Gemini 2.5 Flash & -- & 61.3 & 63.2 & 62.1 \\ \midrule
Qwen2.5-Omni-7B \cite{xu2025qwen25omni} & 7.6B & 46.4 & 46.1 & 46.3 \\
DeSTA 2.5-Audio \cite{desta2025} & 8.8B & 43.3 & 41.7 & 42.6 \\
Qwen2-Audio-Instruct \cite{chu2024qwen2audio} & 8.2B & 30.3 & 27.8 & 29.3 \\
Gemma-3n-E2B-it & 4.4B & 29.6 & 25.8 & 28.0 \\ \midrule
Negative Control (Single-ASR) & 8.8B & 46.8 & 45.8 & 46.4 \\
Qwen2-Audio + TW-Sound580K & 8.2B & 33.0 & 31.8 & 32.5 \\
\textbf{Tai-LALM (Ours)} & 8.8B & \textbf{49.4}  & \textbf{48.8}  & \textbf{49.1} \\ \bottomrule
\end{tabular}
} 
\vspace{-2mm}
\end{table}

Within the compact model regime, Tai-LALM achieves 49.1\% overall accuracy, validating the efficacy of the localized acoustic data in the TW-Sound580K dataset. Fine-tuning on this curated corpus yields a 6.5\% absolute improvement over the DeSTA 2.5-Audio baseline (42.6\% to 49.1\%). Notably, fine-tuning the identical backbone on the raw, unfiltered corpus (Negative Control) establishes a performance floor of 46.4\% due to the dataset's sheer volume . However, residual transcription noise and unaligned soundmarks cap further gains. Employing the VGC pipeline and Dual-ASR arbiter effectively mitigates these artifacts, enabling the model to reach 49.1\% by ensuring high-fidelity cross-modal alignment. This curation also maintains reasoning stability, with accuracy changing only marginally from 49.4\% (single-hop) to 48.8\% (multi-hop) tasks.

Finally, the proposed dataset demonstrates robust transferability, yielding a +3.2\% improvement when applied to the Qwen2-Audio architecture. Fig.~\ref{fig:scaling_law} illustrates a consistent scaling trend (from 5K to 580K pairs), positioning TW-Sound580K as a scalable resource for regional acoustic alignment.
\begin{figure}[htbp]
\centering
\begin{tikzpicture}
\begin{axis}[
    width=0.95\linewidth,
    height=6cm,
    xlabel={Training Data Scale ($\times 10^3$)},
    ylabel={Accuracy (\%)},
    xmin=0, xmax=600,  
    ymin=41, ymax=50.5, 
    xtick={5, 50, 200, 580}, 
    ytick={41, 42, 44, 46, 48, 50}, 
    axis lines=left,
    grid style={dashed, gray!30},
    ymajorgrids=true,
    line width=0.8pt,
    tick label style={font=\small},
    label style={font=\small},
    enlarge x limits={abs=0.2cm},
]

\addplot[dashed, gray, domain=0:600] {42.6}; 
\node[anchor=south east, gray, font=\footnotesize] at (axis cs: 580, 42.6) {Baseline (42.6\%)}; 

\addplot[
    color=blue!70!black,
    mark=*,
    mark size=2.5pt,
    line width=1.5pt,
] coordinates {
    (5, 44.6)
    (50, 44.8)
    (200, 45.2)
    (580, 49.1) 
};

\node[anchor=north, font=\footnotesize, color=black, yshift=-5pt] at (axis cs: 5, 44.6) {44.6};
\node[anchor=south, font=\footnotesize, color=black, yshift=3pt] at (axis cs: 50, 44.8) {44.8};
\node[anchor=south, font=\footnotesize, color=black, yshift=3pt] at (axis cs: 200, 45.2) {45.2};
\node[anchor=south east, font=\footnotesize, color=black, yshift=3pt] at (axis cs: 580, 49.1) {49.1}; 

\end{axis}
\end{tikzpicture}
\caption{Scaling law analysis demonstrating the efficacy of our localized data pipeline on the TW-Sound580K dataset.}
\label{fig:scaling_law}
\end{figure}

\subsection{Ablation Study}
\label{sec:ablation}
\begin{table}[htbp]
\centering
\caption{Ablation study evaluating the impact of data filtering and inference-time ASR injection strategies on TAU Benchmark accuracy.}
\label{tab:ablation_fix}
\resizebox{0.95\linewidth}{!}{
\begin{tabular}{lllc}
\toprule
\textbf{Strategy} & \textbf{SFT Data} & \textbf{ASR Injection} & \textbf{TAU Acc (\%)} \\
\midrule
Zero-shot Baseline & \textemdash & None & 41.1 \\
Direct SFT (Baseline) & Unfiltered Raw TW-580K+ & None & 43.8 \\
\midrule
Single-ASR & Unfiltered Raw TW-580K+ & SenseVoice & 44.5 \\
Single-ASR(Negative Control) & Unfiltered Raw TW-580K+ & Whisper-v3 & 46.4 \\
Dual-ASR & Unfiltered Raw TW-580K+ & Dual-ASR (AC-PPL) & 47.5 \\
\midrule
\textbf{Ours (VGC Pipeline)} & \textbf{Filtered TW-580K} & \textbf{Dual-ASR (AC-PPL)} & \textbf{49.1} \\
\bottomrule
\end{tabular}
}
\vspace{-3mm}
\end{table}

Table \ref{tab:ablation_fix} presents the ablation results across different training and inference configurations. The zero-shot baseline reaches 41.1\% in pure audio mode, which aligns with the 42.6\% default-ASR baseline when text injection is enabled. Fine-tuning on the unfiltered TW-Sound580K+ corpus using only audio features (Direct SFT) improves accuracy to 43.8\%. 

Incorporating inference-time ASR further optimizes performance, yielding 44.5\% with SenseVoice and 46.4\% with Whisper-v3, the latter serving as our primary Negative Control. Switching to Dual-ASR arbitration pushes accuracy to 47.5\%. 

Ultimately, the full pipeline combining VGC-based curation with Dual-ASR arbitration reaches a peak of 49.1\%. Under identical Dual-ASR settings, the 1.6\% gap between unfiltered (47.5\%) and filtered (49.1\%) data confirms that rigorous curation is as critical as inference logic for bridging regional acoustic gaps.

\subsection{Retention of General Capabilities}
\label{sec:retention}

Benchmark results (Table~\ref{tab:general_audio}) indicate that Tai-LALM preserves foundational capabilities during regional adaptation. Notably, LibriSpeech WER improved to 3.92\%, while ESC-50 and CREMA-D exhibited only minor performance trade-offs during regional adaptation.

This stability is rooted in the self-generated target mechanism (Fig. 1-II). By using the frozen backbone LLM to generate target responses from metadata, the fine-tuning objective remains semantically congruent with the original pre-training distribution. This consistency prevents catastrophic forgetting, allowing the model to internalize localized acoustic features within its existing representational space.

\begin{table}[htbp]
\centering
\caption{General audio evaluation. Tai-LALM improves LibriSpeech ASR and maintains competitive non-speech performance, confirming no catastrophic forgetting after localized fine-tuning.}
\label{tab:general_audio}
\resizebox{0.95\linewidth}{!}{
\begin{tabular}{lccc}
\toprule
\textbf{Model} & \textbf{LibriSpeech} & \textbf{ESC-50} & \textbf{CREMA-D} \\
 & (WER\% $\downarrow$) & (Acc\% $\uparrow$) & (Acc\% $\uparrow$) \\
\midrule
DeSTA 2.5-Audio (Zero-shot) & 4.71 & \textbf{34.50} & \textbf{75.22} \\
Tai-LALM (Ours) & \textbf{3.92} & 33.50 & 72.28 \\
\bottomrule
\end{tabular}
}
\vspace{-3mm}
\end{table}


\section{Discussion and Limitations}

Our results indicate that aligning LALMs to regional acoustics is primarily a data-centric challenge. Architectural scaling alone is insufficient for robust sound-to-meaning grounding without localized acoustic semantics. TW-Sound580K addresses this by providing region-specific pairs, enabling models to internalize local soundmarks rather than processing them as noise. The improvements over zero-shot baselines and raw-data fine-tuning emphasize the role of dataset quality. The VGC pipeline and Dual-ASR arbitration facilitate this by filtering ungrounded transcriptions and stabilizing inference. This underscores the utility of regionally situated benchmarks like TAU \cite{tau2025benchmark} in identifying failure modes that are less apparent in globally aligned corpora.

Beyond the Taiwanese context, this pipeline offers a method for regional adaptation. Constructing a localized dataset and applying VGC curation provides a computationally viable alternative to continual pre-training. However, transferring this pipeline to other languages requires calibrating the ASR-injection and routing criteria to match the specific encoder-LLM interface.

\textbf{Limitations:} (1) The VGC curation relies on an empirical threshold ($\tau$); adapting this to new regions requires recalibrating the balance between acoustic diversity and data purity. (2) Dual-ASR arbitration introduces latency and VRAM overhead, necessitating optimization for edge deployment. (3) While Table~\ref{tab:general_audio} verifies the retention of foundational skills, localized performance is primarily evaluated on TAU, which remains the primary benchmark for Taiwanese acoustics.

\section{Conclusion}
\label{sec:conclusion}

This work presents TW-Sound580K and Tai-LALM, which achieves a peak accuracy of 49.1\% on the TAU benchmark, outperforming the Qwen2.5-Omni baseline by 2.8\%. The performance gains underscore the necessity of the VGC pipeline for robust training-time curation and Dual-ASR arbitration for stabilizing inference. By prioritizing high-fidelity data alignment over architectural scaling, this data-centric approach effectively bridges regional acoustic gaps while maintaining the model’s foundational auditory capabilities.

Our future work will prioritize scaling this pipeline to other under-resourced linguistic regions. Beyond literal acoustic-to-text mapping, we aim to investigate the intricate interplay between regional prosody and socio-cultural intent. Moving from basic acoustic reasoning toward a nuanced grasp of pragmatic nuances in regional dialects will be vital for developing truly localized and culturally-aware audio-language models.

\clearpage

\bibliographystyle{IEEEtran}

\bibliography{references}

\begin{thebibliography}{10}
\providecommand{\url}[1]{#1}
\csname url@samestyle\endcsname
\providecommand{\newblock}{\relax}
\providecommand{\bibinfo}[2]{#2}
\providecommand{\BIBentrySTDinterwordspacing}{\spaceskip=0pt\relax}
\providecommand{\BIBentryALTinterwordstretchfactor}{4}
\providecommand{\BIBentryALTinterwordspacing}{\spaceskip=\fontdimen2\font plus
\BIBentryALTinterwordstretchfactor\fontdimen3\font minus \fontdimen4\font\relax}
\providecommand{\BIBforeignlanguage}[2]{{%
\expandafter\ifx\csname l@#1\endcsname\relax
\typeout{** WARNING: IEEEtran.bst: No hyphenation pattern has been}%
\typeout{** loaded for the language `#1'. Using the pattern for}%
\typeout{** the default language instead.}%
\else
\language=\csname l@#1\endcsname
\fi
#2}}
\providecommand{\BIBdecl}{\relax}
\BIBdecl

\bibitem{rubenstein2023audiopalm}
P.~K. Rubenstein, C.~Asawaroengchai, D.~D. Nguyen, A.~Bapna, Z.~Borsos, F.~de~Chaumont~Quitry, P.~Chen, D.~E. Badawy \emph{et~al.}, ``{AudioPaLM}: A large language model that can speak and listen,'' \emph{arXiv preprint arXiv:2306.12925}, 2023.

\bibitem{huang2025dynamicsuperb}
C.-y. Huang, W.-C. Chen, S.-w. Yang, A.~T. Liu, C.-A. Li, Y.-X. Lin, W.-C. Tseng, A.~Diwan \emph{et~al.}, ``{Dynamic-SUPERB Phase-2}: A collaboratively expanding benchmark for measuring the capabilities of spoken language models with 180 tasks,'' in \emph{International Conference on Learning Representations (ICLR)}, 2025.

\bibitem{gong2023ltu}
Y.~Gong, H.~Luo, A.~H. Liu, L.~Karlinsky, and J.~Glass, ``Listen, think, and understand,'' in \emph{International Conference on Learning Representations (ICLR)}, 2024.

\bibitem{tang2023salmonn}
C.~Tang, W.~Yu, G.~Sun, X.~Chen, T.~Tan, W.~Li, L.~Lu, Z.~Ma \emph{et~al.}, ``{SALMONN}: Towards generic hearing abilities for large language models,'' in \emph{International Conference on Learning Representations (ICLR)}, 2024.

\bibitem{chu2023qwenaudio}
Y.~Chu, J.~Xu, X.~Zhou, Q.~Yang, S.~Zhang, Z.~Yan, C.~Zhou, and J.~Zhou, ``{Qwen-Audio}: Advancing universal audio understanding via unified large-scale audio-language models,'' \emph{arXiv preprint arXiv:2311.07919}, 2023.

\bibitem{cao2024culturellm}
C.~Li, M.~Chen, J.~Wang, S.~Sitaram, and X.~Xie, ``{CultureLLM}: Incorporating cultural differences into large language models,'' in \emph{Advances in Neural Information Processing Systems (NeurIPS)}, 2024.

\bibitem{shor2022universal}
J.~Shor, A.~Jansen, W.~Han, D.~Park, and Y.~Zhang, ``Universal paralinguistic speech representations using self-supervised conformers,'' in \emph{IEEE International Conference on Acoustics, Speech and Signal Processing (ICASSP)}, 2022.

\bibitem{yang2024taiwanese}
C.-K. Yang, Y.-K. Fu, C.-A. Li, Y.-C. Lin, Y.-X. Lin, W.-C. Chen, H.~L. Chung, C.-Y. Kuan \emph{et~al.}, ``Building a {Taiwanese Mandarin} spoken language model: A first attempt,'' \emph{arXiv preprint arXiv:2411.07111}, 2024.

\bibitem{kuan2024understanding}
C.-Y. Kuan, W.-P. Huang, and H.-y. Lee, ``Understanding sounds, missing the questions: The challenge of object hallucination in large audio-language models,'' in \emph{Interspeech}, 2024.

\bibitem{lin2025mitigating}
Y.-C. Lin, H.-C. Chou, and H.-y. Lee, ``Mitigating subgroup disparities in multi-label speech emotion recognition: A pseudo-labeling and unsupervised learning approach,'' in \emph{Interspeech}, 2025.

\bibitem{gemmeke2017audioset}
J.~F. Gemmeke, D.~P.~W. Ellis, D.~Freedman, A.~Jansen, W.~Lawrence, R.~C. Moore, M.~Plakal, and M.~Ritter, ``{AudioSet}: An ontology and human-labeled dataset for audio events,'' in \emph{IEEE International Conference on Acoustics, Speech and Signal Processing (ICASSP)}, 2017.

\bibitem{panayotov2015librispeech}
V.~Panayotov, G.~Chen, D.~Povey, and S.~Khudanpur, ``{LibriSpeech}: An {ASR} corpus based on public domain audio books,'' in \emph{IEEE International Conference on Acoustics, Speech and Signal Processing (ICASSP)}, 2015.

\bibitem{zhang2022wenetspeech}
B.~Zhang, H.~Lv, P.~Guo, Q.~Shao, C.~Yang, L.~Xie, X.~Xu, H.~Bu, X.~Chen, C.~Zeng, D.~Wu, and Z.~Peng, ``{WenetSpeech}: A 10000+ hours multi-domain mandarin corpus for speech recognition,'' in \emph{IEEE International Conference on Acoustics, Speech and Signal Processing (ICASSP)}, 2022.

\bibitem{kreuk2023audiogen}
F.~Kreuk, G.~Synnaeve, A.~Polyak, U.~Singer, A.~D{\'e}fossez, J.~Copet, D.~Parikh, Y.~Taigman, and Y.~Adi, ``{AudioGen}: Textually guided audio generation,'' in \emph{International Conference on Learning Representations (ICLR)}, 2023.

\bibitem{chu2025when}
C.~Wang, G.~Deng, X.~Yang, H.~Qiu, and T.~Zhang, ``When audio and text disagree: Revealing text bias in large audio-language models,'' \emph{arXiv preprint arXiv:2508.15407}, 2025.

\bibitem{kim2025wow}
J.~Kim, H.~Yun, S.~H. Woo, C.-H.~H. Yang, and G.~Kim, ``{WoW-Bench}: Evaluating fine-grained acoustic perception in audio-language models via marine mammal vocalizations,'' \emph{arXiv preprint arXiv:2508.20976}, 2025.

\bibitem{nie2022kespeech}
Z.~Tang, D.~Wang, Y.~Xu, J.~Sun, X.~Lei, S.~Zhao, C.~Wen, X.~Tan, C.~Xie, S.~Zhou, R.~Yan, C.~Lv, Y.~Han, W.~Zou, and X.~Li, ``{KeSpeech}: An open source speech dataset of {Mandarin} and its eight subdialects,'' in \emph{Advances in Neural Information Processing Systems (NeurIPS) Datasets and Benchmarks Track}, 2021.

\bibitem{ding2025less}
W.~Ding and F.~Qian, ``{LESS}: Large language model enhanced semi-supervised learning for speech foundational models using in-the-wild data,'' in \emph{IEEE International Conference on Acoustics, Speech and Signal Processing (ICASSP)}, 2026.

\bibitem{ouyang2022training}
L.~Ouyang, J.~Wu, X.~Jiang, D.~Almeida, C.~L. Wainwright, P.~Mishkin, C.~Zhang, S.~Agarwal \emph{et~al.}, ``Training language models to follow instructions with human feedback,'' in \emph{Advances in Neural Information Processing Systems (NeurIPS)}, 2022.

\bibitem{udandarao2025spe-langy}
V.~Udandarao, Z.~Lu, X.~Chang, Y.~Wang, V.~Z. Yao, A.~M. Jose, F.~Faghri, J.~Gardner \emph{et~al.}, ``Data-centric lessons to improve speech-language pretraining,'' \emph{arXiv preprint arXiv:2510.20860}, 2025.

\bibitem{hsu2025reducing}
T.-w. Hsu, K.-H. Lu, C.-H. Chiang, and H.-y. Lee, ``Reducing object hallucination in large audio-language models via audio-aware decoding,'' in \emph{IEEE Automatic Speech Recognition and Understanding Workshop (ASRU)}, 2025.

\bibitem{beyond2025asr}
N.~Glazer, Y.~Segal-Feldman, H.~Segev, A.~Shamsian, A.~Buchnick, G.~Hetz, E.~Fetaya, J.~Keshet \emph{et~al.}, ``Beyond transcription: Mechanistic interpretability in {ASR},'' in \emph{Proceedings of the AAAI Conference on Artificial Intelligence}, 2026.

\bibitem{tau2025benchmark}
Y.-C. Lin, Y.-H. Chen, J.-K. Dong, Y.-H. Huang, S.-C. Chen, Y.-C. Chen, C.-Y. Chen, Y.-J. Lin, Y.-L. Chen, Z.-Y. Chen, I.-N. Tsai, H.-H. Wang, H.-L. Chung, K.-H. Lu, and H.-y. Lee, ``{TAU}: A benchmark for cultural sound understanding beyond semantics,'' in \emph{Proceedings of the IEEE International Conference on Acoustics, Speech and Signal Processing (ICASSP)}, 2026.

\bibitem{desta2025}
K.-H. Lu, Z.~Chen, S.-W. Fu, C.-H.~H. Yang, S.-F. Huang, C.-K. Yang, C.-E. Yu \emph{et~al.}, ``{DeSTA2.5-Audio}: Toward general-purpose large audio language model with self-generated cross-modal alignment,'' \emph{IEEE Transactions on Audio, Speech and Language Processing}, 2026.

\bibitem{sperber2019attention}
M.~Sperber, G.~Neubig, J.~Niehues, and A.~Waibel, ``Attention-passing models for robust and data-efficient end-to-end speech translation,'' \emph{Transactions of the Association for Computational Linguistics (TACL)}, 2019.

\bibitem{atwany2025lost}
H.~Atwany, A.~Waheed, R.~Singh, M.~Choudhury, and B.~Raj, ``Lost in transcription, found in distribution shift: Demystifying hallucination in speech foundation models,'' in \emph{Findings of the Association for Computational Linguistics: ACL}, 2025.

\bibitem{kuan2025teaching}
C.-Y. Kuan and H.-y. Lee, ``Teaching audio-aware large language models what does not hear: Mitigating hallucinations through synthesized negative samples,'' in \emph{Interspeech}, 2025.

\bibitem{frieske2024hallucinations}
R.~Frieske and B.~E. Shi, ``Hallucinations in neural automatic speech recognition: Identifying errors and hallucinatory models,'' \emph{arXiv preprint arXiv:2401.01572}, 2024.

\bibitem{radford2023whisper}
A.~Radford, J.~W. Kim, T.~Xu, G.~Brockman, C.~McLeavey, and I.~Sutskever, ``Robust speech recognition via large-scale weak supervision,'' in \emph{Proceedings of the 40th International Conference on Machine Learning (ICML)}, 2023.

\bibitem{sensevoice2024}
K.~An, Q.~Chen, C.~Deng, Z.~Du, C.~Gao, Z.~Gao, Y.~Gu, T.~He \emph{et~al.}, ``{FunAudioLLM}: Voice understanding and generation foundation models for natural interaction between humans and {LLMs},'' \emph{arXiv preprint arXiv:2407.04051}, 2024.

\bibitem{gemini25_pro2025}
G.~Comanici, E.~Bieber, M.~Schaekermann, I.~Pasupat, N.~Sachdeva, I.~Dhillon, M.~Blistein, O.~Ram \emph{et~al.}, ``{Gemini 2.5}: Pushing the frontier with advanced reasoning, multimodality, long context, and next generation agentic capabilities,'' \emph{arXiv preprint arXiv:2507.06261}, 2025.

\bibitem{xu2025qwen25omni}
J.~Xu, Z.~Guo, J.~He, H.~Hu, T.~He, S.~Bai, K.~Chen, J.~Wang \emph{et~al.}, ``{Qwen2.5-Omni} technical report,'' \emph{arXiv preprint arXiv:2503.20215}, 2025.

\bibitem{chu2024qwen2audio}
Y.~Chu, J.~Xu, Q.~Yang, H.~Wei, X.~Wei, Z.~Guo, Y.~Leng, Y.~Lv, J.~He, J.~Lin, C.~Zhou, and J.~Zhou, ``{Qwen2-Audio} technical report,'' \emph{arXiv preprint arXiv:2407.10759}, 2024.

\end{thebibliography}
\end{document}